\documentclass[12pt,nofootinbib]{article}

\usepackage{slashed}
\usepackage{amsmath,amssymb,graphicx,multicol} 
\usepackage{epsf,color}
\usepackage{cite}
\usepackage{cancel}
\usepackage{framed}

\newcommand{\Tr}{\text{Tr}}
\newcommand{\ifb}{\text{fb}^{-1}}

\newcommand{\nc}{\newcommand}
\nc{\postscript}[2]{\setlength{\epsfxsize}{#2\hsize}\centerline{\epsfbox{#1}}}

\newcommand\beq{\begin{eqnarray}}
\newcommand\eeq{\end{eqnarray}}
\def\bea{\begin{eqnarray}}
\def\eea{\end{eqnarray}}

\def\bit{\begin{itemize}}
\def\eit{\end{itemize}}

\def\baa{\begin{array}}
\def\eaa{\end{array}}

\def\lag{\mathcal L}

\def\simgt{\mathrel{\lower2.5pt\vbox{\lineskip=0pt\baselineskip=0pt
           \hbox{$>$}\hbox{$\sim$}}}}
\def\simlt{\mathrel{\lower2.5pt\vbox{\lineskip=0pt\baselineskip=0pt
           \hbox{$<$}\hbox{$\sim$}}}}

\begin{document}
\begin{titlepage}
\begin{flushright}
\end{flushright}

\vskip.5cm

\begin{center}
{\huge \bf  Model-Independent Bounds \\[0.5cm] on a Light Higgs}
\end{center}
\vskip1cm

\renewcommand{\thefootnote}{\fnsymbol{footnote}}
\begin{center}
{\bf Aleksandr Azatov, Roberto Contino, Jamison Galloway\footnote{email:  aleksandr.azatov@roma1.infn.it, roberto.contino@roma1.infn.it, jamison.galloway@roma1.infn.it}
}
\end{center}
\renewcommand{\thefootnote}{\arabic{footnote}}

\begin{center}
{\it Dipartimento di Fisica, Universit\`a di Roma ``La Sapienza'' \\
and INFN Sezione di Roma, I-00185 Rome, Italy} \\
\vspace*{0.3cm}
\end{center}

\vglue 0.3truecm

\begin{abstract}
\vskip 3pt \noindent
We present up-to-date constraints on a generic Higgs parameter space.  An accurate assessment of these exclusions must take into account statistical, and potentially signal, fluctuations in the data currently taken at the LHC.  For this, we have constructed a straightforward statistical  method for making full use of the data that is publicly available.  We show that, using the expected and observed exclusions which are quoted for each search channel, we can fully reconstruct likelihood profiles under very reasonable and simple assumptions.  Even working with this somewhat limited information, we show that our method is sufficiently accurate to warrant its study and advocate its use over more naive prescriptions. 
Using this method, we can begin to narrow in on the remaining viable parameter space for a Higgs-like scalar state, and to ascertain the  nature of any hints of new physics---Higgs or otherwise---appearing in the data.
\end{abstract}

\end{titlepage}



\section{Introduction}
\label{sec:Intro} \setcounter{equation}{0} \setcounter{footnote}{0}

The  search for a Higgs boson at the LHC has  entered an exciting phase. 
There have been recent excesses of events recorded in various channels by the ATLAS and CMS collaborations 
for a Higgs mass $m_h \approx 125\,$GeV, in the region preferred by the  electroweak (EW) precision tests performed at LEP.
Despite the common view point which
considers the  Higgs as the last missing piece of the successful Standard Model (SM) construction,
the exploration of the TeV scale  that has started at the LHC should be seen rather as
our first mapping of unknown territory, where the theory sector responsible for the breaking of  electroweak (EW) symmetry and the origin of mass
is being tested for the first time. 

Crucial, though indirect, information is  encoded in the LEP precision tests, such as a clear indication  that 
the electroweak symmetry breaking (EWSB) dynamics must possess an approximate custodial symmetry, so as to ensure small corrections 
to the $\rho$ parameter.
If one assumes that the  contribution of the Higgs to the  EW parameters dominates over that of possible additional new states,
LEP suggests that the Higgs must be light and that its coupling to the $W$ and $Z$ vector bosons is within $\sim 15\%$ its SM value.
Even under these assumptions, however, there is no indication from LEP on the value of the couplings of the Higgs to fermions.

On the theoretical side, it is well known that all the successes of the $SU(2)_L \times U(1)_Y$ theory of the EW interactions hold---with  the exception of the LEP precision tests just mentioned---even in absence of a Higgs boson.  The theory can in fact be
formulated in a fully  consistent way by using the  formalism of chiral Lagrangians, which is the standard framework to model
effective field theories with spontaneously broken symmetries.  Such a  description becomes strongly coupled at the scale 
$\Lambda \approx 1-3\,$TeV unless additional states, for example a light Higgs boson, appear below that energy threshold.
In this regard the Higgs model of the SM represents a very peculiar UV completion of the EW chiral Lagrangian, where just one extra 
scalar field is added to the spectrum of known particles with  couplings exactly tuned to ensure perturbativity up to Planckian scales.
While perturbativity implies calculability of the theory, the price to pay is that of an instability of the Higgs mass term against 
radiative corrections, which makes a light elementary Higgs boson highly unnatural.

The fine-tuning problem of the Higgs model is resolved in theories where the Higgs boson is a composite state of new strong dynamics
at the TeV scale \cite{compositeHiggs} or where an additional symmetry, like supersymmetry, protects its mass. 
In a generic theoretical framework, the couplings of the Higgs boson can  differ significantly from their SM values as the result of mixing
with other light scalars or as implied by the composite nature of the Higgs. Given our current limited information on the dynamics responsible 
for breaking EW symmetry, it is  important to keep a general perspective when looking for the Higgs boson at the  colliders.
The EW chiral Lagrangian, with the addition of a light Higgs-like scalar, represents the theoretical 
starting point to 
analyze and optimize the  Higgs searches 
in a model-independent way.

In this work we will show how such a model-independent analysis, once applied to the data collected so far at the colliders,  
can lead to further insight on the Higgs searches, and can perhaps suggest further optimization of the present experimental strategies.
Although a thorough interpretation of the current data would require more detailed information than the one
currently made
public by the experimental collaborations, we have designed an approximate method to extract the likelihood of a given channel using
the expected and observed exclusion limits for the SM Higgs. Such a technique becomes rigorous in the gaussian limit of large number
of counts and turns out to be  accurate under several independent checks that we have performed.
Knowledge of the likelihoods allows one to reinterpret the individual limits in a generic Higgs model and then recombine different 
searches in a  rigorous way. In this regard our method improves on different strategies where the various limits on the Higgs are
individually considered~\cite{HiggsBounds}, or more empirical recipes like a quadrature combination of the limits are adopted.

Our work shares features with previous studies on  model-independent approaches to the extraction of the 
Higgs couplings,  for example the pioneering 
paper of Duhrssen~\cite{duhrssen} and those in Refs.~\cite{Duhrssen:2004cv,Lafaye:2009vr,Bonnet:2011yx}.
Even more similar in the spirit to the present work is the study of the Higgs couplings performed by Refs.~\cite{grojean,Bock:2010nz}
in the  context of composite Higgs theories.

The paper is organized as follows. In section~\ref{sec:lagrangian} we discuss the EW chiral Lagrangian which describes a light Higgs-like
scalar including the complete set of 4-derivative operators which modify the couplings of the Higgs to the vector bosons.
Section~\ref{sec:Stats} is devoted to defining our technique of extracting the likelihoods from existing exclusion limits on the Higgs and discussing
its accuracy. Readers not interested in the details on the method can skip this part and move to section~\ref{sec:limits}, where we apply it to
estimate the model-independent limits on the Higgs couplings and on the strong scale of two benchmark composite Higgs models.
In Section~\ref{sec:125excess} we perform a best fit for the point at $m_h=125\,$GeV, assuming the excess of events observed by CMS and ATLAS is
due to the Higgs. We conclude in section~\ref{sec:Conclusions}.

\section{General Lagrangian for a light Higgs-like scalar}
\label{sec:lagrangian}

Let us consider the case in which a light neutral scalar $h$ exists in addition to the known matter and gauge fields.
The most general description of such Higgs-like particle is obtained by considering 
the EW chiral Lagrangian and adding all possible interactions involving $h$~\cite{Contino:2010mh}.
By requiring an approximate custodial symmetry,
the longitudinal $W$ and $Z$ polarizations correspond to the Nambu-Goldstone (NG) bosons of a global coset 
$SU(2)_L\times SU(2)_R/SU(2)_V$ and can be described by the $2\times 2$ matrix
\begin{equation}
\Sigma(x) = \exp\left( i \sigma^a \chi^a(x)/v \right)\, ,
\end{equation}
where $\sigma^a$ are the Pauli matrices and $v = 246\,$GeV.
The scalar $h$ is assumed to be a singlet of the custodial $SU(2)_V$.
The Lagrangian thus reads:
\begin{equation}
\label{eq:genL}
{\cal L} =   - V(h)  + {\cal L}^{(2)} +  {\cal L}^{(4)} + \dots
\end{equation}
where ${\cal L}^{(n)}$ includes the terms with $n$ derivatives and $V(h)$ is the potential for $h$.
At the level of two derivatives one has~\cite{Contino:2010mh}~\footnote{We omit for simplicity neutrino mass and Yukawa terms,
although they can be included in a straightforward way.}
\begin{equation}
\begin{split}
{\cal L}^{(2)} =
&  \, \frac{1}{2} (\partial_\mu h)^2  + \frac{v^2}{4} \Tr\left( D_\mu \Sigma^\dagger D^\mu \Sigma \right)
    \left( 1+ 2 a\, \frac{h}{v} + b\, \frac{h^2}{v^2} +  \cdots \right)   \\[0.15cm]
& - \frac{v}{\sqrt{2}} \lambda^u_{ij} \; \big( \bar u_L^{(i)} , \bar d_{L}^{(i)} \big) \, \Sigma \,   \big(u^{(i)}_R, 0\big)^T
       \left( 1+ c_u\, \frac{h}{v} + c_{2u} \, \frac{h^2}{v^2} + \cdots \right)+h.c. \\[0.15cm]
& - \frac{v}{\sqrt{2}} \lambda^d_{ij} \; \big( \bar u_L^{(i)} , \bar d_{L}^{(i)} \big)  \, \Sigma \,   \big(0, d^{(i)}_R\big)^T
       \left( 1+ c_d\, \frac{h}{v} + c_{2d} \, \frac{h^2}{v^2} + \cdots \right)+h.c. \\[0.15cm]
& - \frac{v}{\sqrt{2}} \lambda^l_{ij} \; \big( \bar \nu_L^{(i)} , \bar l_{L}^{(i)} \big)  \, \Sigma \,   \big(0, l^{(i)}_R\big)^T
       \left( 1+ c_l\, \frac{h}{v} + c_{2l} \, \frac{h^2}{v^2} + \cdots \right)+h.c. \\[0.15cm]
\end{split}
\end{equation}
where $a$, $b$, $c_{u,d,l}$, $c_{2u,2d,2l}$ are arbitrary dimensionless coefficients, and  $c_{u,d,l}$, $c_{2u,2d,2l}$ have been assumed to be flavor-diagonal to 
avoid inducing dangerous flavor-changing processes. An  implicit sum over flavor indices $i,j = 1,2,3$
has been understood.
Similarly, the potential can also be expanded in powers of $h$, 
\begin{equation}
V(h) = \frac{1}{2} m_h^2 h^2 + d_3\, \frac{1}{6} \left(\frac{3 m_h^2}{v}\right) h^3 + d_4\, \frac{1}{24} \left(\frac{3 m_h^2}{v^2}\right) h^4 + \dots
\end{equation}
where $d_3$, $d_4$ are arbitrary coefficients and $m_h$ is the mass of the scalar $h$.
As discussed in Ref.~\cite{Contino:2010mh}, for generic values of the coefficients the theory is strongly interacting at large energies.
However, for the specific choice $a=b=c_u=c_d=c_e=d_3=d_4=1$ and vanishing higher-order terms, all the scattering amplitudes remain 
perturbative (and unitary)
up to very high energies, provided the scalar $h$ is light. This is indeed the SM limit, in which $h$ is identified with the physical Higgs boson.
For any other choice of  coefficients the exchange of  $h$ only partially cancels the energy growth of the scattering amplitudes of   NG bosons,
and the Lagrangian (\ref{eq:genL}) must be regarded as an effective description valid below some cutoff scale~$\Lambda$.

In this general case, it is still appropriate to refer to $h$ as a Higgs boson if it forms a doublet of $SU(2)_L$ together with the NG bosons $\chi$, 
and as such it plays a role in the breaking of  EW symmetry.  This is naturally realized in   theories of composite Higgs, where $h$ emerges
as a light pseudo-NG boson of a larger dynamically-broken global 
symmetry~\cite{compositeHiggs,Agashe:2004rs,Contino:2006qr,Galloway:2010bp,Giudice:2007fh}.    
The shift symmetry acting on the Higgs in this case allows one to resum all powers of $h$ at a given derivative order. 
At  leading chiral order, this implies that 
 the coefficients in ${\cal L}^{(2)}$ are all functions of $\xi = (v/f)^2$, where $f$ is the decay constant of the composite Higgs.
For small~$\xi$, the effective Lagrangian of such a strongly-interacting light Higgs (SILH) has been fully characterized by Ref.~\cite{Giudice:2007fh}
in terms of a finite number of dimension-6 operators. In particular, it has been shown that $a$ and $b$ follow a universal trajectory in the small $\xi$ limit.

Other scenarios are however possible, in which for example $h$ is a bound state of the dynamics responsible for the breaking of the EW symmetry,
but  does not form an $SU(2)_L$ doublet together with the $\chi$ fields. In fact, it could even well be that $h$ is a Higgs-like impostor, and
plays no role in EWSB. This is for example the case of a light dilaton~\cite{dilaton}.
In all cases, the Lagrangian (\ref{eq:genL}) is a valid effective description for $h$ at energies lower than the cutoff scale.
For convenience, in the following we will refer to $h$ as the Higgs boson even 
for generic values of its couplings.

At the level of four derivatives, it is convenient to write the Lagrangian as a sum of operators $O_i$,
\begin{equation}
{\cal L}^{(4)} = \sum_i O_i\, ,
\end{equation}
whose Higgs dependence is encoded by polynomials 
\begin{equation}
F_i(h) = \alpha_i^{(0)} +  \alpha_i^{(1)} \, h +  \alpha_i^{(2)} \, h^2 + \dots
\end{equation}
with arbitrary  coefficients $\alpha_i^{(n)}$. 
The bosonic operators  that lead to  cubic and quartic vertices of NG bosons and gauge fields with one or two Higgs bosons
are:~\footnote{In a previous version of this paper, the operator 
$\Tr\!\left[ (D_\mu \Sigma)^\dagger (D_\nu \Sigma) \right] \partial^\mu \partial^\nu F(h)$ appeared in Eq.~(\ref{eq:O12}) in place of $O_2$.
Such an operator can however be removed by use of the equations of motion and integration by parts. The operator $O_2$ correctly appeared
in the list of Ref.~\cite{Alonso:2012px}.}
\begin{align}
\label{eq:O12}
\begin{split}
O_{1} & = \Tr\!\left[ (D_\mu \Sigma)^\dagger (D^\mu \Sigma) \right]  (\partial_\nu F_1(h))^2 \\[0.2cm]
O_{2} & = \Tr\!\left[ (D_\mu \Sigma)^\dagger (D_\nu \Sigma) \right] \partial^\mu F_{2}(h) \partial^\nu F_{2^\prime}(h)
\end{split}
\\[0.5cm]
\label{eq:OGG}
\begin{split}
O_{GG} & = G_{\mu\nu} G^{\mu\nu} \, F_{GG}(h) \\[0.2cm]
O_{BB} & = B_{\mu\nu} B^{\mu\nu} \, F_{BB}(h) 
\end{split}
\\[0.5cm]
\begin{split}
O_{W} & = D_\mu W^a_{\mu\nu} \, \Tr\!\left[ \Sigma^\dagger \sigma^a i\overleftrightarrow D_\nu \Sigma \right]  F_W(h) \\[0.2cm]
O_{B} & = -\partial_\mu B_{\mu\nu} \, \Tr\!\left[ \Sigma^\dagger i\overleftrightarrow D_\nu \Sigma \, \sigma^3  \right]  F_B(h) 
\end{split}
\\[0.5cm]
\begin{split}
O_{WH} & = i\, W^a_{\mu\nu} \,\Tr\!\left[ (D^\mu \Sigma)^\dagger \sigma^a D^\nu \Sigma \right] F_{WH}(h) \\[0.2cm]
O_{BH} & = -i\, B_{\mu\nu} \,\Tr\!\left[ (D^\mu \Sigma)^\dagger (D^\nu \Sigma) \sigma^3 \right] F_{BH}(h) 
\end{split}
\\[0.5cm]
\label{eq:OWdH}
\begin{split}
O_{W\partial H} & =  
 \frac{1}{2}\, W^a_{\mu\nu} \,\Tr\!\left[ \Sigma^\dagger \sigma^a i\overleftrightarrow D^\mu \Sigma \right]  \partial^\nu F_{W\partial H}(h) \\[0.2cm]
O_{B\partial H} &=  
 -\frac{1}{2}\, B_{\mu\nu} \,\Tr\!\left[ \Sigma^\dagger  i\overleftrightarrow D^\mu \Sigma \sigma^3 \right]  \partial^\nu F_{W\partial B}(h) \, .
\end{split}
\end{align}
For simplicity, we do not consider fermionic operators in ${\cal L}^{(4)}$. Their effects are suppressed
if the SM fermions couple weakly to the  EWSB  dynamics, in which case the bosonic operators of Eqs.~(\ref{eq:O12})-(\ref{eq:OWdH}) give the main effects.
The assumption of weak fermionic couplings might not be accurate for the top and bottom quarks, see for example the discussion in Ref.~\cite{Giudice:2007fh}.
The operators $O_{GG}$, $O_{BB}$ contribute to the coupling of the Higgs to a pair of gluons and photons and are thus relevant for 
the LHC searches, while $O_{W}$, $O_{B}$ contribute to the $S$ parameter. 
In the case of a composite Higgs, where $h$ is part of an $SU(2)_L$ doublet, at leading order in $\xi$ all the polynomials are fixed to 
the quadratic form $F_i(h) = (1+h/v)^2 (1+ O(h^3) + O(\xi))$, 
and the operators (\ref{eq:OGG})-(\ref{eq:OWdH}) correspond to the SILH Lagrangian.~\footnote{Once written in terms of the $SU(2)_L$ doublet Higgs field,
$O_{1,2}$ correspond instead to dimension-8 subleading operators.}  As pointed out in Ref.~\cite{Giudice:2007fh}, since  $O_{GG}$, $O_{BB}$ do not 
respect the Higgs  shift  symmetry, their coefficient will be suppressed by an extra factor $(\lambda^2/g_\rho^2)$, where $g_\rho$ is the coupling strength
of the strong sector, and $\lambda$ is some (weaker) coupling that breaks explicitly the NG global symmetry. 
For example,  $O_{GG}$, $O_{BB}$ can be generated by the one-loop exchange of vector-like composite fermions~\cite{Low:2009di,Azatov:2011qy}.

The Lagrangian (\ref{eq:genL}) represents the most general (effective) description of a light Higgs under the following  assumptions:
\textit{i)} possible new states are heavy and do not significantly affect the physics below the cutoff scale. In particular, this implies that there are 
no other light states to  which the Higgs can decay;
\textit{ii)} the EWSB dynamics possesses
a custodial symmetry; \textit{iii)} there are no flavor-changing neutral-current processes mediated at tree-level by the Higgs.
While the (at least approximate) validity of the last two assumptions is strongly supported by
the current experimental data, the first assumption  is simply driven by the request
of simplicity, and it can be relaxed by adding to the effective Lagrangian possible new light  states, such as  additional scalars, 
which might be discovered in the future.
For the moment, assuming no such additional light states exist, eq.~(\ref{eq:genL})
 allows for a general parametrization of
the couplings of the Higgs to the  fermions and to the gauge bosons free from (additional) theoretical prejudice, and as such it is the starting point for a 
model-independent interpretation of the experimental searches for a Higgs boson under way at the LHC and Tevatron.

\begin{figure}[tbp]
\begin{center}
\includegraphics[width=0.5\textwidth,clip,angle=0]{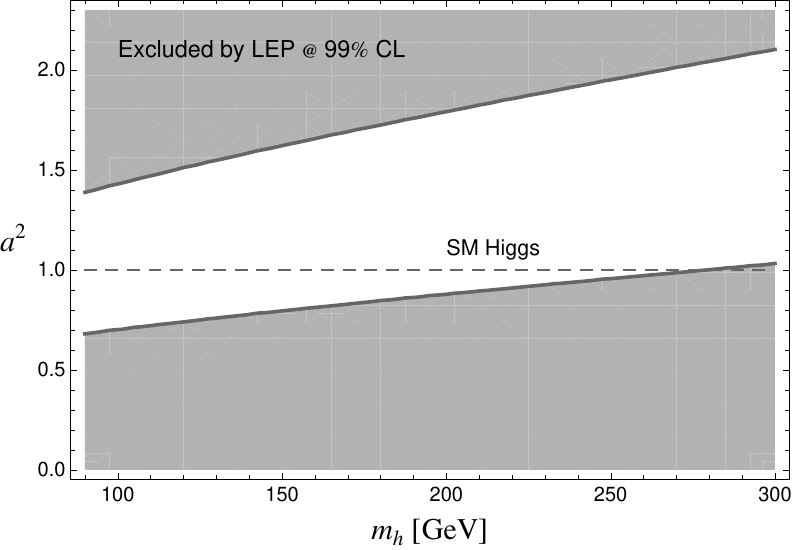}
\caption[]{\small
Limits on the coupling $a^2$ implied by the LEP precision tests for $\Lambda = 4\pi v/\sqrt{1-a^2}$ and $m_t =173.2\,$GeV.
The gray region is excluded at 99\%~CL.
\label{fig:EWPTconstraint}
}
\end{center}
\end{figure}

It is important to notice that with the exception of direct searches, the only experimental information  
is on the coupling of the Higgs to vector bosons:
the  precision tests performed at LEP on the EW observables are sensitive to the Higgs contribution at one loop to the vector boson self energies, 
and thus set a constraint on $a$ 
for a given mass $m_h$. If one compares to the SM case, the additional contribution to the EW parameters $\epsilon_{1,3}$~\footnote{We recall that
$\Delta\epsilon_1 = \Delta \hat T$, $\Delta\epsilon_3 = \Delta \hat S$ where $\hat T$, $\hat S$~\cite{Barbieri:2004qk} are proportional to the 
Peskin-Takeuchi $S$, $T$  parameters~\cite{Peskin:1991sw}.} is
\begin{equation}
\begin{split}
\Delta \epsilon_{1} & =- \frac{3}{16\pi}\, \frac{\alpha(m_Z)}{\cos^2\theta_W} \,  (1-a^2)  \log\left(\frac{\Lambda^2}{m_h^2}\right) \\[0.5cm]
\Delta \epsilon_{3} & =+ \frac{1}{48\pi}\, \frac{\alpha(m_Z)}{\sin^2\theta_W}  \, (1-a^2)  \log\left(\frac{\Lambda^2}{m_h^2}\right) \, .
\end{split}
\end{equation}
Figure~\ref{fig:EWPTconstraint} shows the 99\%CL limits  on $a^2$  obtained by performing a fit to the LEP data with 
$\Lambda = 4\pi v/\sqrt{1-a^2}$.~\footnote{We make use of a $\chi^2$ function of four parameters~\cite{fit}, $\epsilon_{1,2,3,b}$, and set 
$\epsilon_{2}$, $\epsilon_{b}$ to their SM value.} Sizable deviations from the SM value $a=1$ are still allowed; for example, for $m_h = 125\,$GeV one has 
$0.84 \leq a^2 \leq 1.4$.
%
%
It is important to notice that no constraint on the other Higgs couplings (for example $c$ and $b$) follows from the LEP precision tests.
On the other hand, important information on all the single-Higgs couplings follows from the direct searches at LEP, Tevatron, and LHC.

Although in general the experimental data can and should be used to extract \textit{all} the relevant Higgs couplings in (\ref{eq:genL}), 
in this initial survey we will focus on those of a single Higgs to two weak bosons ($a$) and to two SM fermions, and we will set the latter
to be  the same for up and down quarks and for leptons ($c = c_u = c_d =c_l$).
We will thus assume that the effects of the other couplings (for example those from $O_{GG}$ and $O_{BB}$) are subdominant.
This is in fact the case in two simple models of composite Higgs that we will  adopt as useful benchmark theories to illustrate our results.
The first one is the  minimal $SO(5)/SO(4)$ model with SM fermions embedded into spinorial representations of 
$SO(5)$, which has been dubbed MCHM4~\cite{Agashe:2004rs}. In this model 
all single-Higgs couplings are rescaled by a common function of $\xi$,
\begin{equation}
\text{MCHM4:} \hspace{2cm} a = c = \sqrt{1-\xi} \, ,
\end{equation}
so that the Higgs production cross sections get rescaled by a universal factor, whereas the decay branching ratios are not modified compared to their
SM values. The same relations are predicted in the Minimal Conformal Technicolor model~\cite{Galloway:2010bp}.
The second benchmark theory that we will consider is the 
$SO(5)/SO(4)$ MCHM5 model with SM fermions embedded into fundamentals of $SO(5)$~\cite{Contino:2006qr}. It predicts 
a different rescaling of the Higgs couplings to  fermions and vector bosons,
\begin{equation}
\text{MCHM5:} \hspace{2cm} a = \sqrt{1-\xi} \, , \qquad c = \frac{1-2\xi}{\sqrt{1-\xi}}\, ,
\end{equation}
which in turn leads to
a different pattern of decay rates compared to the SM. In particular, for $\xi \to 1/2$ one finds in this theory a concrete realization of the 
possibility of a fermiophobic Higgs. In this limit the theory requires a UV completion at a scale $\Lambda \sim 4\pi f \simeq 4.4\,$TeV.

In the following sections we will show how  the current experimental information from the SM Higgs searches can be used to get an 
accurate estimate of the model-independent constraints that can be set on the 
couplings $a$, $c$ for a given value of the Higgs mass $m_h$. 
By means of the same technique, we will be also able to 
derive the limits on $\xi$ in the benchmark composite Higgs models MCHM4 and MCHM5.

\section{The Statistical Method}
\label{sec:Stats}

The strongest direct constraints on the coefficients $a$, $c$ come from the Higgs searches under way at the LHC.
The results for each decay channel $i$ are expressed in terms of a strength modifier $\mu^i$, 
defined as the signal (Higgs) yield in SM units for a given fixed value of $m_h$~\cite{LHCsummer}:
\begin{equation}
\mu^i = \frac{n^i_s}{(n_s^i)^{SM}} \, .
\end{equation}
If no significant excess of events compared to the background (no Higgs) expectation is observed, a $95\%$~CL limit is set on $\mu$; if instead
an excess is observed, the ATLAS and CMS collaborations report the best fit value of $\mu$ for a given hypothesis on $m_h$. 
In either case the result is derived by constructing a likelihood function $p(n_{obs}| n_{s}+n_b)$ using the signal ($n_s$), background ($n_b$) 
and observed ($n_{obs}$) yields. In the Bayesian approach,~\footnote{Results by the ATLAS and CMS collaborations are derived in two
different statistical methods: the Bayesian method and a hybrid Bayesian-frequentist technique~\cite{LHCsummer}. Although the latter has been chosen as 
the standard technique used to report the collaborations' results, internal derivation of the limits is also performed using the Bayesian framework. 
In this work we will use the Bayesian framework, which  seems to be the simplest and most logical approach for our purposes. See~\cite{D'Agostini:2003nk} for 
a primer.}
a posterior probability density function of $\mu$ is then constructed by assessing some
prior $\pi(\mu)$ on $\mu$:
\begin{equation}
\label{eq:posterior}
p(\mu | n_{obs}) = p(n_{obs} | \mu \, n^{SM}_s + n_b) \times \pi(\mu) \, .
\end{equation}
To derive limits, a flat prior for $\mu\geq 0$ (vanishing for $\mu <0$) is adopted, and the $95\%$~CL limit on~$\mu$ is computed as
that value $\mu_{95\%}$ such that the integral of $p(\mu | n_{obs})$ from $\mu=0$ to $\mu = \mu_{95\%}$ is 0.95.
The result so obtained gives the limit on the (overall) factor by which the SM Higgs yield can be amplified, for a given value $m_h$.
Values $\mu_{95\%} < 1$ thus exclude at $95\%$~CL the SM Higgs for that particular value of the Higgs mass.

For given numbers of expected and observed events, the likelihood is modeled by a Poisson distribution~\footnote{In cases in which an unbinned
likelihood is constructed~\cite{LHCsummer}, use of a binned one is expected to give similar results. See for example the discussion in Section 8 
of~\cite{CMS:2012tw} for the case of $h\to\gamma\gamma$.}
\begin{equation}
\label{eq:likelihood}
p(n_{obs} | \mu \, n^{SM}_s + n_b)  = \frac{1}{n_{obs}!} \, e^{-(\mu\cdot n_s^{SM}+nb)} \left( \mu\cdot n_s^{SM}+n_b \right)^{n_{obs}}\, .
\end{equation}
In a generic theory, for each channel $i$, the signal strength modifier $\mu^i$ can be computed provided one knows the Higgs production cross section
for each production mode $p$, the efficiencies $\zeta^p_i$ of the kinematic cuts, and the Higgs decay branching fraction:
\begin{equation}
\label{eq:mui}
\mu^i \equiv \frac{n_s^i}{(n_s^i)^{SM}} = \frac{\sum_p \sigma_p \times \zeta^p_i}{\sum_p \sigma_p^{SM} \times \zeta^p_i} \times \frac{BR_i}{BR^{SM}_i}\, .
\end{equation}
Notice that the efficiencies of the kinematic cuts depend in general  on the production mode, and are thus crucial to correctly
compute $\mu^i$.
A rigorous assessment of the bounds implied by the Higgs searches on a generic beyond-the-SM (BSM) theory, such as that of eq.~(\ref{eq:genL}), thus 
requires two ingredients:
\begin{enumerate}
\item The likelihood  for each channel $i$ as a function of $\mu$
\item The cut efficiencies $\zeta^p_i$  for each channel $i$ and production mode $p$
\end{enumerate}
Without  this information, it is not possible to derive the exact constraints
on theories different from the SM unless they predict a simple universal rescaling of all the Higgs  cross sections.
Knowledge  of the cut efficiencies  allows one to  derive the bounds implied by each individual channel
on the parameter space of any BSM model.
This is, for example, what the dedicated code \texttt{HiggsBounds}~\cite{HiggsBounds} 
does by considering only those experimental searches where, to good approximation, only one production mode is relevant (as a consequence 
of the kinematic cuts).
In general, however, a consistent statistical combination of the  various channels can be done only by knowing the individual likelihoods.
Unfortunately, \textit{neither the likelihoods nor the cut efficiencies are currently publicly provided by ATLAS and CMS}.~\footnote{The cut efficiencies
are provided only in select cases, e.g. the $\tau \tau$ mode of CMS.  Here, however, the information is available only for one representative value of the Higgs mass, which does not suffice to construct exact likelihoods over the whole mass range, as would be needed to probe the broader parameter space.}

Given the importance of having a broader, model-independent perspective on the Higgs searches, we find it useful to 
try to find possible approaches that can lead to 
an accurate estimate of the bounds on the  couplings in eq.~(\ref{eq:genL}),
by making use of the current information made public by the experimental collaborations.
Below we describe a method that allows one to reconstruct the  likelihood of each channel  given the 
expected and observed 95\%~CL limits on the signal strength modifier, which are the only two numbers that are readily available 
for a given value of $m_h$.
As we will discuss in detail, this method becomes exact in the asymptotic (Gaussian) limit of large event counts, which makes it clearly
preferable over other less rigorous recipes sometimes used to combine the limits.

\subsection{A Technique to Extract the Likelihoods in the Gaussian Limit }

\begin{figure}[tbp]
\begin{center}
\includegraphics[width = 0.45\linewidth]{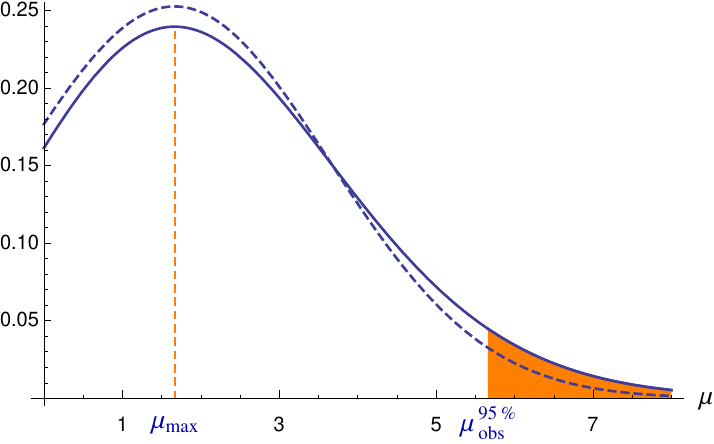} 
\caption{\small
Posterior probability $p(\mu|n_{obs})$ obtained for $n_{obs} = 35$, $n_b=30$, $n_s^{SM}=3$ (continuous curve). 
In this example the maximum is at $\mu_{max} = 5/3$, 
and the 95\%~CL limit on $\mu$ is $\mu^{95\%}_{obs} = 5.66$.  The dashed curve shows the approximating Gaussian 
with mean $\mu_{max}$ and standard deviation $\sigma_{obs} = \sqrt{35}/3$.
}
\label{fig:posterior}
\end{center}
\end{figure}

In general, once considered as a function of $\mu$, the posterior probability (\ref{eq:posterior}) depends on three parameters
($n_s$, $n_b$ and $n_{obs}$), while, as noticed above, we can make use of only two numbers (the expected and observed 95\% CL limits on $\mu$).
However, if the number of observed events is large,  $n_{obs} \gg 1$, the likelihood asymptotically tends to a Gaussian with mean $n_{obs}$ and 
standard deviation 
$\sqrt{n_{obs}}$:~\footnote{Eq.~(\ref{eq:gaussianlimit}) is a special case of the central limit for the Gamma distribution, see for example~\cite{book}.  
When considered as a function of $n$, $p( n_{obs} | n )$ is indeed proportional to a Gamma distribution with shape parameter $k = n+1$ and scale 
parameter $\theta=1$.
Any  factor which does not depend on $n$ can be  dropped, as the overall normalization of the posterior probability will be fixed
at the end. A simple way to prove the asymptotic convergence (\ref{eq:gaussianlimit}) is by considering the  difference  between
$p( n_{obs} | n )$ and the Gaussian at some fixed number of standard deviations away from the maximum: $(n-n_{obs})/\sqrt{n_{obs}} \sim$ a few.
For $n_{obs} \gg 1$ this implies $\Delta = (n - n_{obs})/n_{obs} \ll 1$, so that
\begin{equation}
p( n_{obs} | n ) \propto ( 1 + \Delta )^{n_{obs}} e^{- n_{obs} \Delta} =
\left( 1 - \frac{1}{2} n_{obs} \Delta^2 + O(\Delta^3) \right) = e^{- \Delta^2 n_{obs}/2} + O(\Delta^3)\, , 
\end{equation}
where we made an expansion for small $\Delta$.
}
\begin{equation}
\label{eq:gaussianlimit}
p( n_{obs} | n ) \propto e^{-n} \, n^{n_{obs}} \ \longrightarrow \ e^{- (n-n_{obs})^2/2 n_{obs}}\, .
\end{equation}
In practice, the approximation is already good for $n_{obs} \gtrsim 10$.
In this asymptotic limit the posterior probability (as a function of $\mu$) depends on just two combinations of $n_s$, $n_b$, $n_{obs}$:
\begin{equation}
p(\mu | n_{obs}) \propto e^{  -(\mu - \mu_{max})^2/2\sigma^2_{obs} }   \, , \quad\quad
\mu_{max} = \frac{n_{obs}-n_b}{n_s^{SM}} \, , \quad \sigma_{obs} = \frac{\sqrt{n_{obs}}}{n_s^{SM}}\, .
\end{equation}
The parameter $\mu_{max}$, in particular, determines the location of the maximum of the probability and 
measures by how much  the 
 number of observed events has fluctuated
from the pure background expectation compared
to the number of SM signal events, see Fig.~\ref{fig:posterior}.
As we will now show, the information provided by the experimental collaborations is sufficient, under simple specific assumptions, to determine
$\mu_{max}$, $\sigma_{obs}$ and thus reconstruct the likelihood.
%
%
First,  the value of $\mu_{max}$ and $\sigma_{obs}$ must be such to reproduce the 95\%~CL observed limit on~$\mu$:
\begin{equation}
\label{eq:firstequation}
0.95 = \int \! d\mu \  \, p(\mu | n_{obs})  
  \simeq \frac{ \displaystyle \int^{\mu^{95\%}_{obs}}_0 \! d\mu \ \, e^{  \textstyle -\frac{(\mu - \mu_{max})^2}{2\sigma^2_{obs} }} }%
                      {\displaystyle \int^{\infty}_0 \! d\mu \ \, e^{\textstyle  -\frac{(\mu - \mu_{max})^2}{2\sigma^2_{obs} }} }  
  = \frac{\text{Erf}\left( \frac{\mu^{95\%}_{obs} - \mu_{\max}}{\sqrt{2} \sigma_{obs}}\right) + \text{Erf}\left( \frac{\mu_{\max}}{\sqrt{2} \sigma_{obs}}\right)}%
                                      {1+ \text{Erf}\left( \frac{\mu_{\max}}{\sqrt{2} \sigma_{obs}}\right)}\, .
\end{equation}
A second relation is  obtained from the expected  95\% CL  limit, which is derived as above but  setting $n_{obs} = n_b$ (pure background 
hypothesis). In this case the posterior probability $p(\mu | n_{obs} = n_b)$ is approximated in the asymptotic limit by a Gaussian with zero mean and 
standard deviation $\sigma_{exp} = \sqrt{n_{b}}/n_s^{SM}$,
as one can see by setting $n_{obs} = n_b$ in eq.~(\ref{eq:gaussianlimit}).
The relation implied by the   95\% CL expected limit  is:
\begin{equation}
0.95 = \int \! d\mu \  \, p(\mu | n_{obs} = n_b)  \simeq \sqrt{\frac{2}{\pi \sigma_{exp}^2}} 
 \int^{\mu^{95\%}_{exp}}_0 \! d\mu \ \, e^{  -\mu ^2/2\sigma^2_{exp} } 
 = \text{Erf}\left( \frac{\mu^{95\%}_{exp}}{\sqrt{2} \sigma_{exp}}\right)\, ,
\end{equation}
which admits the simple solution:
\begin{equation}
\label{eq:simplesolution}
\frac{\sqrt{n_{b}}}{n_s^{SM}} = \sigma_{exp} = \frac{\mu^{95\%}_{exp}}{1.96}\, .
\end{equation}
Although this is not an equation on the parameters of the posterior $p(\mu | n_{obs})$, it
can be used to determine $\sigma_{obs}$ provided
 the fluctuation is small compared to the number of background events:
\begin{equation} 
\label{eq:condition}
\frac{n_{obs} - n_b}{n_b} \ll 1\, .
\end{equation}
Notice that if $n_s \ll  n_b$ the fluctuation can still be large compared to the number of \textit{signal} events, that is, $\mu_{max} \sim O(1)$.
If eq.~(\ref{eq:condition}) is satisfied, one can approximate $\sigma_{obs} \simeq \sigma_{exp} = \sqrt{n_{b}}/n_s^{SM}$ and extract $\mu_{max}$ by 
numerically  solving eq.~(\ref{eq:firstequation}). In this way the likelihood is fully reconstructed as a function of $\mu$. By using 
eq.~(\ref{eq:mui}) one can then evaluate the value of $\mu$ in terms
of the parameters of any generic Higgs model, and thus obtain the likelihood as a function of these parameters. 
Finally, the combined bound from several channels is obtained by multiplying their  likelihoods.

At this point a comment is in order regarding the validity of combining the limits from  individual channels in quadrature, which is what  has sometimes 
 been used in the literature to estimate the constraints implied by the Higgs searches on generic BSM models.
It is simple to see (and well known) that the combination in quadrature is justified, in the gaussian limit, for the {\it expected} limits. It just follows from
the simple fact that the product of  gaussians with zero mean and standard deviations $\sigma_{exp}^i$ is still a gaussian with zero mean 
and variance $(\sigma_{exp}^{comb})^2 = 1/\sum_i (1/\sigma_{exp}^{i})^2$. Applying eq.~(\ref{eq:simplesolution}) to each channel then leads to the inverse 
quadrature formula:
\begin{equation}
\label{eq:quadrature}
\mu^{95\%}_{comb, exp} = \frac{1}{\displaystyle \sqrt{\sum_i \frac{1}{(\mu^{95\%}_{i, exp})^2}}}\, .
\end{equation}
On the other hand, this formula can\textit{not} be used to combine the {\it observed} limits, since in that case the combined limit
obtained by means of the product of likelihoods cannot be  expressed simply in terms of the individual limits.
Using eq.~(\ref{eq:quadrature}) for the observed limits  does not properly take into account the experimental fluctuations.
A quantitative comparison between the naive quadrature combination and our method is reported in 
Figs.~\ref{fig:BayesComb},~\ref{fig:errors},~\ref{fig:MCHM4} and discussed below.

So far we have tacitly neglected  possible systematic errors on the number of signal and background events. In the Bayesian approach 
they are simply incorporated by marginalizing the posterior probability over a set of nuisance parameters,  taking into account possible 
correlations~\cite{LHCsummer}. 
In order to show how our method accounts for such systematic  effects, we consider for simplicity only two nuisance parameters,  
$\theta_s$, $\theta_b$, which reflect the overall systematic uncertainty respectively on the number of signal and background  events. 
The posterior probability in this case is given by
\begin{equation}
\label{eq:marginalizedposterior}
p(\mu | n_{obs}) \propto 
\int_{-\infty}^{+\infty} \!\!\! d\theta_b \int_{-\infty}^{+\infty} \!\!\! d\theta_s
\ \ p(n_{obs} | \mu\cdot n_s^{SM}\,  e^{\theta_s k_s} + n_b\, e^{\theta_b k_b} )\,  e^{-\theta_b^2/2}\, e^{-\theta_s^2/2}
\end{equation}
where $k_{s} = \Delta_{s}/n_{s}^{SM}$, $k_{b} = \Delta_{b}/n_{b}$ and $\Delta_{s}$ ($\Delta_{b}$) is the systematic error on the number of signal 
(background) events.
The nuisance parameters have been assumed to be distributed with  LogNormal pdfs, as commonly done by  CMS and ATLAS 
to ensure that the number of signal and background events never becomes negative.
However, if the  systematic errors are small, $\Delta_{b}/n_{b}, \Delta_{s}/n_s^{SM} \ll 1$, the LogNormal distributions can be approximated
by (truncated) Gaussians.~\footnote{Truncation of the integral at $\theta_{s} = -n_{s}^{SM}/\Delta_{s}$, $\theta_{b} = -n_{b}/\Delta_{b}$ is required to 
avoid having a negative number  of events.}  In this case one obtains (up to an overall normalization)
\begin{equation}
p(\mu | n_{obs} ) \simeq \frac{\displaystyle 
e^{\textstyle - \frac{(\mu\, n^{SM}_s + n_b - n_{obs})^2}{2 (n_{obs} + \Delta_b^2 + \mu^2 \Delta_s^2)} }}{\sqrt{2\pi ( n_{obs} + \Delta_b^2 + \mu^2 \Delta_s^2 ) }} \, .
\end{equation}
Although this not a Gaussian function of $\mu$, in many practical cases one can neglect the dependence on $\mu$ 
in the denominator of the exponent and in the overall factor. 
The resulting probability can then be approximated by a Gaussian with mean $\mu_{max}$ and
modified standard deviation $\sigma_{obs} = \sqrt{n_{obs} + \Delta_b^2 }/n_s^{SM}$. Similarly, the expected posterior probability,
$p(\mu| n_{obs} = n_b)$, is approximately a Gaussian with zero mean and modified standard deviation 
$\sigma_{exp} = \sqrt{n_{b} + \Delta_b^2}/n_s^{SM}$. 
The exact condition for this gaussian approximation to hold is
\begin{equation}
\label{eq:limitforsystematics}
\frac{\Delta_s}{n_s^{SM}} \, \frac{n_{obs}-n_b}{\sqrt{\Delta_b^2 + n_{obs}}} \ll 1 \, .
\end{equation}
If eqs.~(\ref{eq:condition}) and (\ref{eq:limitforsystematics}) are satisfied, then our  method to extract the likelihood from the expected and observed 95\%~CL limits
can  be applied, the only modification with respect to the previous discussion is that now the parameters $\sigma_{obs}$, $\sigma_{exp}$
get a contribution also from the systematic error on the number of background events.~\footnote{In fact, approximating 
$\sigma_{obs} \simeq \sigma_{exp}$, as required in our method to extract the likelihood,  is even more accurate if $\Delta_b^2/n_b$ is not small, 
while $\Delta_b/n_b \ll 1$.}
As a final comment we notice that the size of the 68\% and 95\%  bands reported for the expected exclusion limit  by CMS and ATLAS 
(green and yellow bands) gives in principle some additional information on how the limit changes when the nuisance parameters vary.
Since however  such information does not seem easy to use for reconstructing the likelihoods, we have not considered it.

It is useful to summarize the conditions on which our method relies:
\begin{enumerate}
\item The number of observed events must be large (Gaussian limit).
\item The fluctuations must be small compared to the number of background events, though not necessarily small compared 
to the number  of signal events: condition (\ref{eq:condition}).
\item The systematic error on the number of background events must be  small, $\Delta_{b}/n_{b} \ll 1$, and that on the number of signal events
must be negligible: $\Delta_{s}/n_s^{SM} \ll 1$ plus condition~(\ref{eq:limitforsystematics}).
\end{enumerate}

\subsection{Discussion of the accuracy of our method}

Before applying it to derive the model-independent bounds on the couplings $a$, $c$,
we want to discuss here the accuracy of our method for extracting the likelihoods. 
A first test of its validity comes from the comparison with the official limit on $\mu$ obtained by combining all the searches
performed by a single LHC experiment.
We find that the combined bound derived using our technique
reproduces with good accuracy the official curve in the whole range of Higgs masses.

Figure~\ref{fig:BayesComb} shows the comparison
for CMS using the full 2011 data set ($4.6 - 4.8\,\ifb$)~\cite{CMS:combo}.
%
\begin{figure}[tb]
\begin{center}
\includegraphics[height=6.6cm]{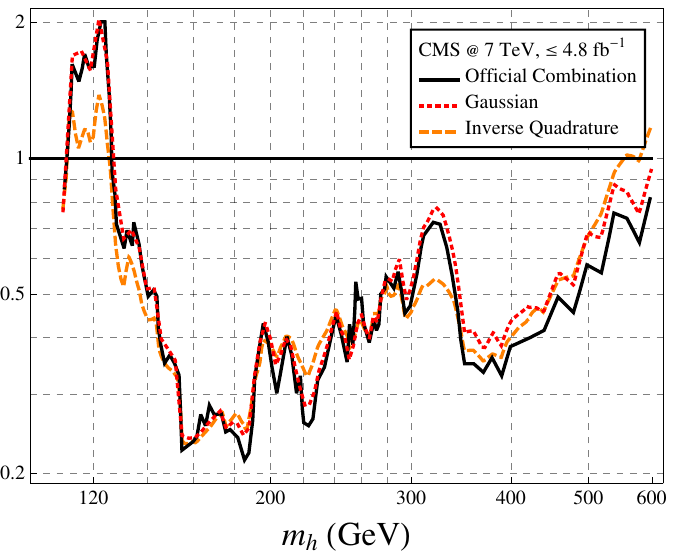} 
\includegraphics[height=6.6cm]{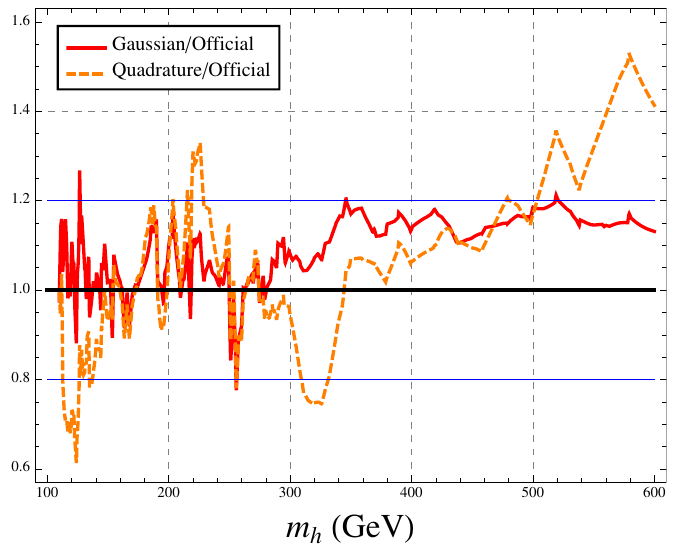} 
\caption{\small
Left panel:  95\%~CL observed limits on $\mu$ obtained by combining all CMS searches with  different techniques: the 
continuous black curve is the official CMS limit, the dotted red and dashed orange curves are obtained respectively with our method and
by a naive  quadrature combination.
Right panel: relative deviation of the limits obtained with these two latter approaches from the official combination. 
The blue band at $\pm 20\%$ is for illustration.
}
\label{fig:BayesComb}
\end{center}
\end{figure} 
%
When available, 
in fact only for $h\to WW$, we have used the  limits from each of the subchannels of a given search to reconstruct their individual
likelihoods. For those searches where only a combined  limit 
was available, like for $h\to\gamma\gamma$, 
we have used that to reconstruct the overall likelihood. Although in most of the cases we could find only 95\%~CL limits obtained with the
CLs frequentist method, we did make use of the Bayesian limits in those few cases where they were available. On the other hand, the two approaches
have been shown to lead to very similar results (see for example~\cite{CMS:comboPAS}), so that we expect
that  using  CLs  limits instead of  Bayesian ones leads to a difference in our results which is
within the error of the gaussian approximation.

As shown in the right plot of Fig.~\ref{fig:BayesComb}, the relative difference 
between the 95\%~CL  limits obtained with our Gaussian technique and the official CMS curve is always smaller than~20\%, 
and in fact our combination typically  errs on the conservative side. 
For the sake of comparison, we show also the result of adding observed exclusions in inverse quadrature as an approximation
of the  total.  As expected, we find that this approach is incapable of accounting for competing fluctuations in different channels, and can lead to 
regions of unrealistically strong exclusions.

A more detailed comparison is possible by focusing on the $h\to WW\to l\nu l\nu$ channel. Two different kinds of analysis are performed in this case 
by CMS: the first makes use of a boosted decision tree technique, the second is purely cut-based. For the latter analysis, the number of signal, 
background,  and observed events is made publicly available at  $m_h = 120\ {\rm GeV}$ for each of the five categories 
considered~\cite{Chatrchyan:2012ty,WWtwiki},  
which makes it possible to fully construct the individual likelihoods using eq.~(\ref{eq:marginalizedposterior}). We find that these constructed likelihoods 
are able to reproduce the  median  95\%~CL expected and observed official limits on $\mu$ within $15-20\%$.
This shows that (at least for this channel) a simple two-dimensional marginalization, 
eq.~(\ref{eq:marginalizedposterior}), captures the most important effects of the systematic uncertainties.  
Figure~\ref{fig:errors} shows the relative difference between these constructed likelihoods and those extracted with our method from the published 
95\%~CL limits as a function of $\mu$, for the representative point $m_h = 120\ {\rm GeV}$. 
%
\begin{figure}[tbp]
\begin{center}
\includegraphics[width = 0.65\textwidth]{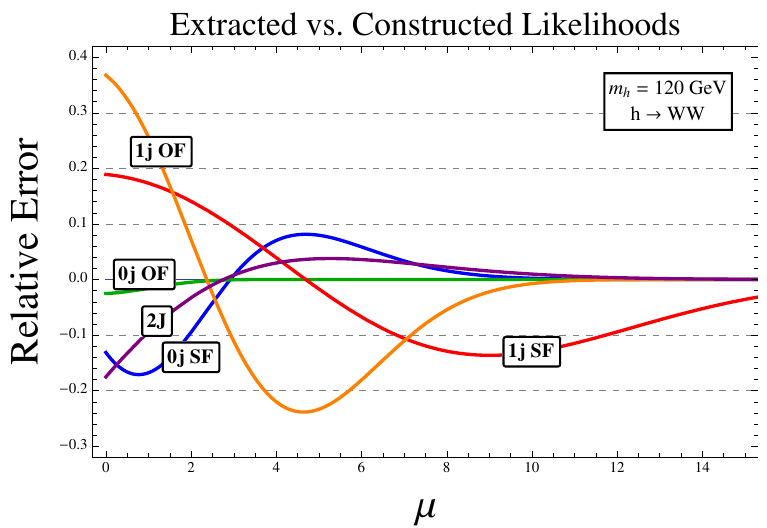} 
\caption{\small
Relative error between extracted and constructed likelihoods for the five  $h \to WW$ categories of CMS, as a function of the signal strength 
modifier $\mu$. In each case the extracted Gaussian likelihood is found to approximate the one constructed  from event numbers
typically to within~$20\%$.}
\label{fig:errors}
\end{center}
\end{figure} 
%
For convenience, we report in Table~\ref{table:WW120} 
the number of events in each channel that we have used, as given by the CMS collaboration~\cite{WWtwiki}.
%
\begin{table}[tbp]
\begin{center}
\begin{tabular}{c | c | c | c | c | c }
Final State: jets/leptons & $n_B$ & $\Delta n_B$ & $n_S$ & $\Delta n_S$ & $n_{\rm obs}$  \cr   \hline \hline
0-jet, Same Flavor & 50.6 & 9.8 & 4.7 & 1.1 & 49\\
0-jet, Opp. Flavor & 86.1 & 8.2 & 11.0 & 2.5 & 87 \\
1-jet, Same Flavor & 20.4 & 2.6 & 1.7 & 0.5 & 26\\
1-jet, Opp. Flavor  & 39.1 & 5.3 & 4.8 & 1.7 & 46 \\
2-jet & 11.3 & 3.6 & 1.1 & 0.1 & 8  \\ \hline
\end{tabular}
\caption{\small
Background, signal, and observed events (with related uncertainties) reported by CMS in the five $WW$ categories  
for $m_h = 120 \ {\rm GeV}$, $\int\! dt\, \lag \leq 4.7 \, {\rm fb}^{-1}$~\cite{WWtwiki}.}
\label{table:WW120}
\end{center}
\label{default}
\end{table}
%
With the exception of the 1-jetOF category, where the agreement is slightly worse, 
the extracted likelihood is seen to be accurate at the level of  $\pm 20 \%$.
The precision of our method is also clearly illustrated by Fig.~\ref{fig:WWacplane}, which shows the observed 95\%~CL exclusion curve in the plane
$(a,c)$ as obtained from the combination of the five $WW$ categories by using our method (orange curve) and by using the likelihoods constructed
from the event numbers of Table~\ref{table:WW120} (blue area).
%
\begin{figure}[tb]
\begin{center}
\includegraphics[width = 0.45\linewidth]{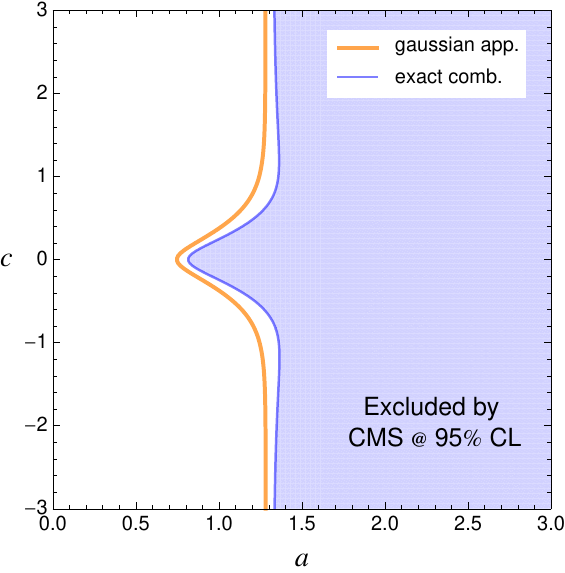} 
\caption{\small
95\%~CL observed limits in the plane $(a,c)$ obtained by combining the five $WW$ categories in CMS for $m_h = 120\,$GeV.
The blue and orange curves are obtained using respectively the likelihoods constructed from the  number of events in Table~\ref{table:WW120} 
(exact combination) and the  likelihoods reconstructed with our method (gaussian approximation).
}
\label{fig:WWacplane}
\end{center}
\end{figure} 
%
In either case we rescaled the 2-jet category assuming that its yield entirely comes from the VBF Higgs production, as a consequence
of the cuts imposed. The other four categories are instead rescaled by assuming that they are entirely dominated by the gluon-fusion production.
While this is clearly a rough approximation, it should be sufficiently accurate in most of the $(a,c)$ plane and  conservative in the  fermiophobic
region $c\sim0$.
The agreement between the two exclusion curves in Fig.~\ref{fig:WWacplane} is good over the whole $c$ range. 
The stronger exclusion around $c\sim 0$ is a consequence of the greater significance of the VBF channel in this limit. As we will discuss 
in sec.~\ref{sec:125excess}, the inclusive analysis performed by ATLAS for $h\to WW$ is much less sensitive to the fermiophobic region.

To summarize, the above results show that our method works accurately enough and can thus be used to derive a robust estimate of the bounds
implied by the LHC searches on a generic Higgs model.

\section{Model independent bounds}
\label{sec:limits} 

In this section we apply our  method 
to derive the model-independent limits on the couplings $a$, $c$
in the framework of the  effective Lagrangian~(\ref{eq:genL}). We will also show the bounds on $\xi$ in the case of the 
two benchmark composite Higgs models MCHM4 and MCHM5.
All the plots have been derived making use of the CMS results obtained through the analysis of the full 
2011 data set ($4.6 - 4.8\,\ifb$)~\cite{CMS:combo}. Similar conclusions are also obtained using the ATLAS data. 
We will not show the exclusions implied by  Tevatron searches as they turn out to be weaker than the LHC ones.
As  mentioned in the previous section, we reconstructed the likelihoods of individual subchannels in a given search whenever possible.
In each case the signal strength modifier has been computed as a function of $(a,c)$  by taking into account the exclusive or inclusive nature
of the search. In particular, we assumed that the signal yield is fully dominated by  the  associated Higgs production in $h\to b\bar b$,
by VBF production for the 2-jet category of $h\to WW$, and by gluon fusion production for the 0-jet and 1-jet categories of $h\to WW$.
All the other searches ($h\to ZZ$, $h\to\tau\tau$, $h\to \gamma\gamma$) have been considered as inclusive.
Since for these channels the cut efficiencies $\zeta^p_i$ of eq.~(\ref{eq:mui}) are not provided by CMS, we have assumed them 
to be constant (\textit{i.e.} independent of the Higgs production mechanism), although this is known to be a somewhat inaccurate approximation, especially 
in the limit $|c| \ll 1$  where the gluon fusion cross section is suppressed  compared to its SM value. 
The same assumption was made in the previous studies  of Ref.~\cite{grojean}.

We begin with the MCHM4 model, where the Higgs production cross sections are rescaled by a common factor. The same results apply 
to any  model with universal rescaling, as is the case for example in minimal conformal TC.
In this case the 95\%~CL limits on $\xi$ are simply obtained from those on the signal strength modifier by setting $\mu = 1-\xi$.
The result is shown in Fig.~\ref{fig:MCHM4}, where we report the curves obtained by means of the official CMS limit, our gaussian method,
and the  inverse quadrature  combination.  The curve obtained with the latter method agrees with the results of Ref.~\cite{grojean}.
We have superimposed also the 
region selected by the LEP precision data at 99\%~CL, which 
has been obtained, as in Fig.~\ref{fig:EWPTconstraint}, by considering just the Higgs contribution to the EW observables.
Since the contribution of additional states, naturally present in composite Higgs theories, can give an important contribution to the EW observables,
this region should be considered simply as  indicative rather than as a sharp exclusion contour.
%
\begin{figure}[tb]
\begin{center}
\includegraphics[width=0.6\linewidth]{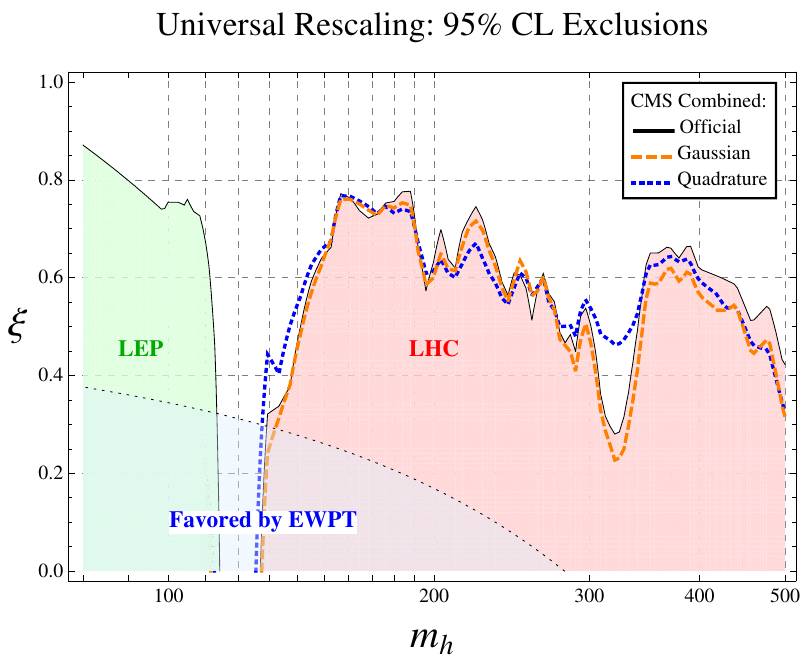} 
\caption{\small
Current 95\% CL exclusion limits on models with $a=c=\sqrt{1-\xi}$.  The  region excluded by  LHC (LEP) data is 
shown in light red (green).  We show here the comparison of the three different combination prescriptions discussed in the text: 
the solid black line corresponds to the official CMS combination in the CL$_S$ asymptotic approach, the dashed orange line is obtained  
using our gaussian method, and the dotted blue line shows the result of combination in quadrature.}
\label{fig:MCHM4}
\end{center}
\end{figure} 
%
We see that values $\xi \gtrsim 0.5 -0.6$, which correspond to a suppression  $g_{higgs}/g_{higgs}^{SM} \lesssim 0.5$ in the Higgs couplings,
are needed for a heavy Higgs to escape the current LHC exclusion. In the case of a light composite Higgs and small $\xi$, on the other hand, the
allowed range of $m_h$ is roughly the same as for a SM Higgs.

The current exclusion limits on $\xi$ for the  MCHM5 are shown in Fig.~\ref{fig:MCHM5}. 
%
\begin{figure}[tbp]
\begin{center}
\includegraphics[width=0.6\textwidth]{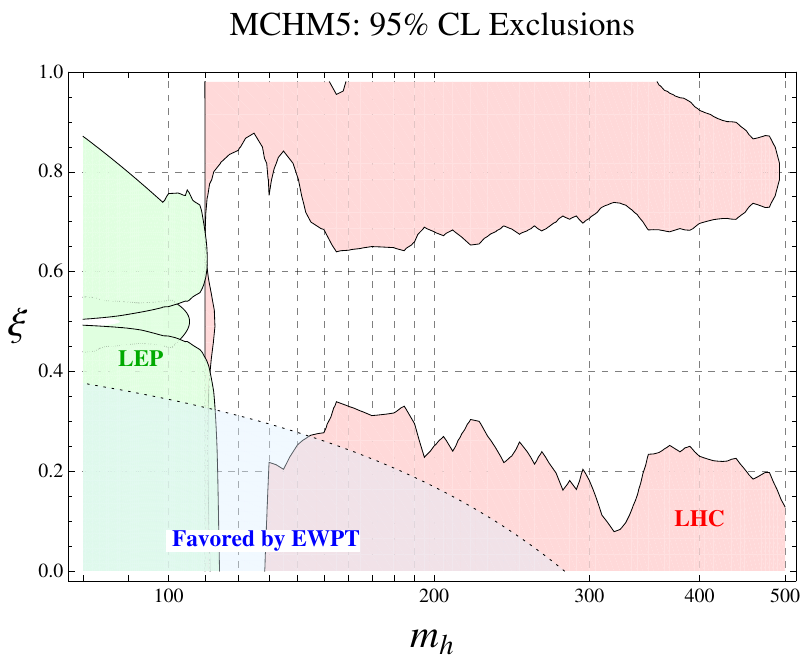} 
\caption{\small
Current 95\% CL exclusion limits on $\xi$ in the MCHM5 ($a=\sqrt{1-\xi}$, $c=(1-2\xi)/\sqrt{1-\xi}$) as obtained with our method. 
The  region excluded by LHC (LEP) data is  shown in light red~(green). }
\label{fig:MCHM5}
\end{center}
\end{figure} 
%
As previously discussed, in this model the region  $\xi \sim 1/2$ corresponds to a limit 
where the Higgs is fermiophobic, and its production rate is suppressed. This implies that a heavy Higgs can escape the current limits in an ample range
of values $\xi \sim 0.3 - 0.7$.  A similar plot has been derived in Ref.~\cite{grojean} by combining limits in inverse quadrature.

Finally, we report in Fig.~\ref{fig:ac95} the current limits on the plane $(a,c)$ for some reference values of $m_h$.
%
\begin{figure}[tbp]
\begin{center}
\includegraphics[width=0.55\textwidth]{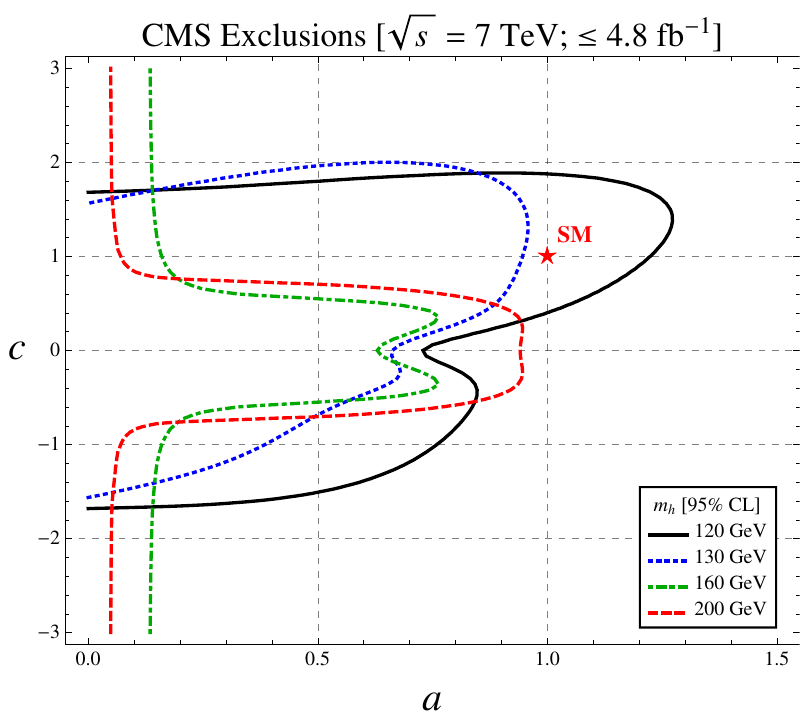} 
\caption{\small
Current exclusions in the plane $(a,c)$ for various Higgs masses as obtained with our method: 
the area to the right of each curve is excluded at 95\% CL.  These exclusions combine all search channels at CMS, with the full 2011
data set  $\int \! dt\, \lag \leq 4.8 \, {\rm fb}^{-1}$.}
\label{fig:ac95}
\end{center}
\end{figure} 
%
They have been obtained by combining all the CMS search channels using our method. Note that the likelihoods are now treated as fully two-dimensional functions $p(a,c|n_{obs})$, with production and branching ratio rescaling factors themselves functions of $a$ and $c$.  This implies a difference of priors relative to the results of Figs.~\ref{fig:MCHM4} and \ref{fig:MCHM5}, where the two couplings were mapped to a single overall rescaling, $\mu$, whose prior is assumed to be flat over the interval $[0,\infty)$.  The two-dimensional exclusions can thus be constructed simply by determining isocontours enclosing a desired fraction of the normalized likelihood.  For this case, we assume priors that are flat over the range $0\leq a \leq 3$ and $-3 \leq c \leq 3$, and zero elsewhere.

We notice that for $m_h = 120,\, 130\,$GeV
the exclusion curve is sensitive to the relative sign between $a$ and $c$, while for heavier Higgs masses the curves 
are symmetric under $c\to -c$. This is due to the importance for light $m_h$ played by the $\gamma\gamma$ channel, the only one sensitive to the
relative sign through the decay width to two photons.
In particular, for negative $c/a$ the interference between the one-loop top and $W$ contributions to the  decay width
is constructive and the constraint is  stronger.

\section{The 125\,GeV Excess}
\label{sec:125excess} 

A somewhat anomalous point has emerged in both CMS and ATLAS at $m_h \approx 125\, {\rm GeV}$, with surpluses of events being registered in 
multiple channels by both experiments.  Although the statistical significance in each case is below $3 \sigma$ once look-elsewhere effects are included, 
it is certainly interesting to consider the shape of the total likelihood in this neighborhood.   We show the result of this exercise in Fig.~\ref{fig:125},
for $m_h = 125\,$GeV.  
%
\begin{figure}[tb]
\begin{center}
\includegraphics[width=0.47\linewidth]{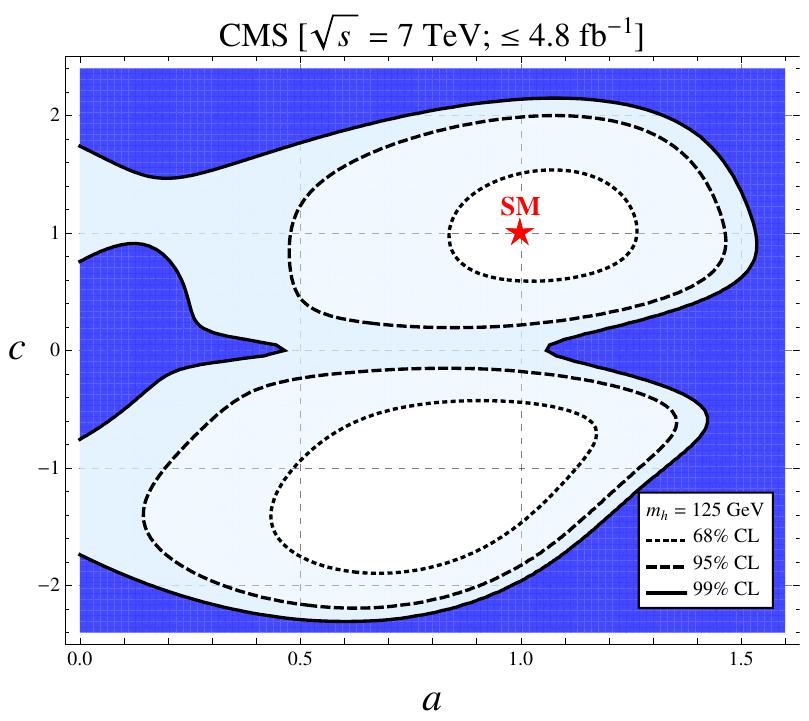} 
\hspace{0.5cm}
\includegraphics[width=0.47\linewidth]{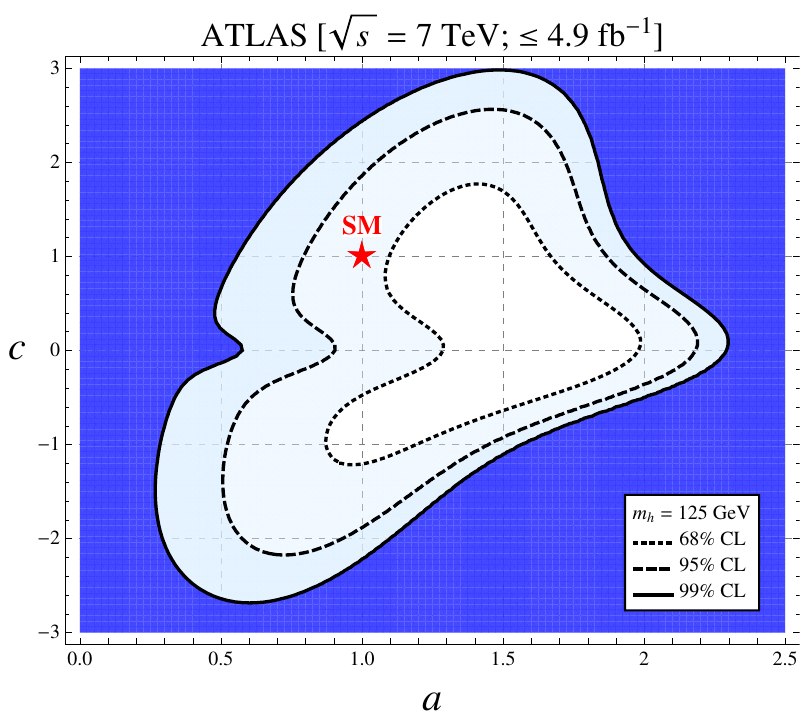} 
\caption{\small
Isocontours of 68\%, 95\% and 99\%  probability in the plane $(a,c)$  
for a 125 GeV Higgs coming from CMS (left) and ATLAS (right).  
In each case the posterior probability has been constructed using the method described in sec.~\ref{sec:Stats}.
}
\label{fig:125}
\end{center}
\end{figure} 

The plot on the left shows the best fit in the plane $(a,c)$ obtained with our method  using the CMS data 
($\int \! dt\, \lag \leq 4.8 \, {\rm fb}^{-1}$)~\cite{CMS:combo}.
The posterior probability has two peaks, which indicate two  solutions preferred by the current data.
The first maximum is for $(a \simeq 0.9, c \simeq -1.2)$ and has the  highest probability. It corresponds to a solution for $(a,c)$ that
leads to an enhanced yield in $\gamma\gamma$ and a slight suppression in $WW$, $ZZ$ compared to the SM expectation.
It is useful to define the ratio
\begin{equation}
R_i \equiv \frac{\sigma \times BR(i)}{[\sigma \times BR(i)]_{SM}}\, ,
\end{equation}
where $\sigma$ stands for the Higgs total  production cross section  (\textit{i.e.} summed over all production modes), which indicates
the change in the signal yield compared to its SM value for an inclusive search in the channel $i$.
For  $(a = 0.9, c = -1.2)$ one has  $R_{\gamma\gamma}\simeq 2.3$ and $R_{WW} = R_{ZZ} \simeq 0.86$.
The enhancement in $\gamma\gamma$ follows from the constructive interference in the relative decay width,
$\Gamma(\gamma\gamma) \propto |1.8\, c - 8.3\, a|^2$, that arises for negative $c$.
An enhanced yield in $\gamma\gamma$ and a slight suppression of $WW$, $ZZ$ is in fact exactly what the best fit of the individual channels
performed by CMS also points to (see Fig.~4 of Ref.~\cite{CMS:combo}). 
We thus find that such a pattern of rates can be easily reproduced for $c \sim -1$, which ensures an enhanced $\gamma\gamma$ while
predicting a gluon fusion production cross section close to its SM value. 
The second maximum of the probability is for $(a \simeq 1.15, c \simeq 1.0)$. It is smaller than the first peak, as the shorter
isocontours indicate. This solution roughly corresponds to the combined best fit of CMS where all rates are $20\% - 30\%$ larger than their 
SM expectations ($R_{\gamma\gamma}\simeq 1.4$ and $R_{WW} = R_{ZZ} \simeq 1.3$ for $(a = 1.15, c = 1.0)$).
While the  maximum at $c\simeq 1$ already emerges from the fit  when including  the channels $WW$, $ZZ$ and $\gamma\gamma$ alone, we find that
the $\tau\tau$ search plays an important role in shaping the highest peak and excluding  points with large and negative $c$.

The plot on the right of Fig.~\ref{fig:125} shows the best fit in the plane $(a,c)$ obtained using the full 2011 ATLAS data set 
($\int \! dt\, \lag \leq 4.9 \, {\rm fb}^{-1}$)~\cite{ATLAS:combo}.
Compared to the corresponding analysis of CMS, the sensitivity of the $h\to WW$ inclusive search in ATLAS (in which the 2-jet VBF category 
is not singled out) is much
weaker in the fermiophobic region $c\sim 0$. This implies a much broader region where the posterior probability  is large, instead of two
disconnected smaller islands. Furthermore, the excess in the $ZZ$ channel seen by ATLAS leads to a best fit for $(a\simeq 1.5, c\simeq 0.45)$,
which corresponds to $R_{\gamma\gamma}\simeq 2.0$, $R_{WW} = R_{ZZ} \simeq 1.4$. Notice that in this case the enhancement of the $\gamma\gamma$
rate, as well as that of $WW$ and $ZZ$, follows from  $a > 1$.  In fact, this can be obtained only in specific UV completions of the
effective Lagrangian~(\ref{eq:genL}), see Refs.~\cite{Low:2009di,Falkowski:2012vh}. If confirmed, it would thus be a strong hint on the nature and the role 
of the Higgs.  On the other hand, another way to obtain an enhanced rate in all channels except $b\bar b$ is that of  suppressing the total Higgs 
decay width by having $c_b < 1$.~\footnote{We thank Riccardo Rattazzi for drawing our attention to this possibility. See also~\cite{rattazzi}
for a discussion.}
This solution is not accessible in our 2-dimensional fit where all the fermion couplings were constrained to be the same, but can be naturally realized in particular models: for example, Ref.~\cite{Azatov:2011qy} demonstrates such a possibility in composite models, while the models of Ref.~\cite{SCTC} allow for such a solution in a supersymmetric setting at large $\tan \beta$ and Refs.~\cite{Akeroyd:1998ui,Ferreira:2011aa} discuss the more general implications for two Higgs doublet models.  As such, having $c_b<1$ represents
a simple possibility that should be clearly considered when analyzing the data.

Although these preliminary indications from ATLAS and CMS do not yet fit into a coherent picture, it is clear that a simple
analysis of the data in terms of the  parameters $a$,~$c$ will represent an important and powerful tool to determine the nature of the 
Higgs boson, should the hints of its presence at $125\,$GeV be confirmed.
In this regard, we consider  it useful to provide the plot of Fig.~\ref{fig:Risocontours}, which shows the isocurves of constant $R_{\gamma\gamma}$ and
$R_{WW}=R_{ZZ}$ in the plane $(a,c)$ for $m_h = 125\,$GeV. 
%
\begin{figure}[tb]
\begin{center}
\includegraphics[width=0.47\linewidth]{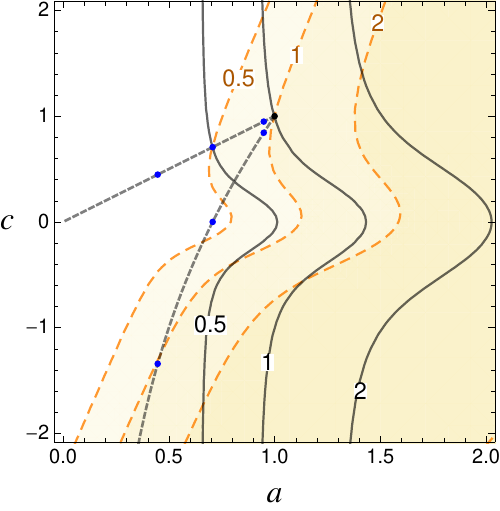} 
\caption{\small
Isocontours with $R_{\gamma\gamma} = 0.5,1,2$ (orange long dashed curves) and $R_{WW}=R_{ZZ}=0.5,1,2$ (continuous back curves) in the plane $(a,c)$
for $m_h = 125\,$GeV. The upper (lower) short dashed gray curve is the  trajectory predicted in the MCHM4 (MCHM5). The blue dots show the points 
with $\xi = 0.1, 0.5, 0.8$.
}
\label{fig:Risocontours}
\end{center}
\end{figure} 
%
The different solutions preferred by CMS and ATLAS can be easily recognized along the isocurve 
$R_{\gamma\gamma}\sim 2$. These solutions cannot be reached by following the trajectories predicted in the composite
models MCHM4 and MCHM5 (shown in the plot as short dashed gray curves).
In the MCHM5, in particular, there cannot be an enhancement in the yield of an inclusive $\gamma\gamma$ search.
Although in the fermiophobic limit $\xi\to 1/2$  the branching fraction to $\gamma\gamma$ gets enhanced by up to a factor 7,
this is more than compensated by the drop in the gluon fusion cross section. At the same time, however, the yield in the VBF subchannel
of an exclusive $\gamma\gamma$ search can be enhanced by up to a factor 3 for  $\xi\sim 1/2$. 

The possibility that the enhanced yield in $\gamma\gamma$  might be due to a fermiophobic Higgs has been recently 
suggested by Ref.~\cite{Gabrielli:2012yz}. The main support to this idea comes from the latest exclusive analysis of  $\gamma\gamma$  performed
by CMS~\cite{CMS:2012tw}, which in fact reports a larger excess in the VBF category than in the other four dominated by gluon fusion production.
Our global  fit of the CMS data in Fig.~\ref{fig:125}, however, seems to disfavor the fermiophobic solution ($a=1$, $c=0$). As already mentioned,
a dominant role for $c\sim 0$ is played by the exclusive  analysis of $h\to WW$~\cite{Chatrchyan:2012ty}.
Indeed, for $m_{h} = 125\,$GeV the fermiophobic
solution ($a=1$, $c=0$) implies a strong enhancement in the branching ratio of not just the $\gamma\gamma$ channel, but of $WW$ as well
(respectively a factor $\sim 6.6$ in $BR(\gamma\gamma)$  and  4.1 in $BR(WW)$). For an inclusive $WW$ search such an increase is more than 
compensated by a decrease in the gluon fusion production cross section, but this is not the case for a category dominated by events produced
through the VBF process. 
The absence of a substantial excess in the 2-jet category of the  $WW$ analysis of CMS is in fact what 
disfavors  a fermiophobic Higgs more strongly  in the current data.~\footnote{In fact, for both $m_h =120\,$GeV and $130\,$GeV the 2-jet category
has a depletion in the number of observed events compared to the pure background expectation.}

This simple example shows how much more powerful it can be to perform an exclusive analysis instead of  an inclusive one when it comes
to extracting information of the Higgs couplings. This is especially true for the $\gamma\gamma$ channel~\cite{us}, but also   for $WW$ as seen above;  we expect for the same to be true for $\tau\tau$ as well. This observation is in fact one of the main points put forward by the authors 
of Ref.~\cite{Gabrielli:2012yz}. In this regard we must notice that the published information in~\cite{CMS:2012tw} was not sufficient
to include the $\gamma\gamma$ channel in an exclusive fashion in our fit (only the combined limit over all categories is given in~\cite{CMS:2012tw}).
At the best fit point  ($a=0.9$, $c=-1.2$) selected by our fit, we find that the signal yield in a VBF-dominated subchannel
(like the 2-jet category of the CMS analysis) is enhanced by a factor $R^{VBF}_{\gamma\gamma} = 1.4$, compared to $R_{\gamma\gamma} = 2.3$ of the inclusive 
yield. As previously  noticed, the best fit of the individual categories done by CMS prefers a larger enhancement in the 2-jet subchannel. 
This pattern can in fact be easily reproduced for $c$ negative and smaller than $a$ in magnitude. For example, the point ($a = 1$, $c = -0.8$)
implies $R^{VBF}_{\gamma\gamma} = 3.1$, $R_{\gamma\gamma} = 2.1$.  We thus expect that  once a fully exclusive inclusion of the $\gamma\gamma$
channel into the fit is performed, the  region of maximum probability with $c <0$ will shrink and the location of the maximum will migrate
to smaller values of $|c|$.

\section{Conclusions}
\label{sec:Conclusions} 

The majority of the searches for the Higgs boson at the LHC and Tevatron are optimized  for the SM Higgs and results
are reported accordingly. However, it is of extreme importance to have a broader perspective on the nature of the Higgs boson,
especially since the origin of the EW symmetry breaking remains very uncertain.
In this work we have shown how a model-independent analysis on the Higgs couplings can be performed already with the current data, and should
be carried out in future analyses. The theoretical foundation is that of the EW chiral Lagrangian in eq.~(\ref{eq:genL}), which relies 
on three simple assumptions:  \textit{i)} a  Higgs-like scalar is the only new light particle in the spectrum, and  additional  states
are much heavier and do not  significantly affect the Higgs phenomenology at low energy;  \textit{ii)}  the dynamics that breaks the EW symmetry 
possesses an approximate custodial symmetry;  \textit{iii)} no dangerous tree-level FCNC are mediated by the exchange of the Higgs boson.
If needed, the first assumption can be relaxed and additional states can be consistently added to the  Lagrangian by following
the rules of the chiral expansion.

Depending on the value of the Higgs couplings in eq.~(\ref{eq:genL}), the phenomenology that follows can be quite different from that of the SM Higgs.
Although eventually one would like to perform a completely general analysis and individually measure as many Higgs couplings
as possible, in this work we have considered a simplified though interesting scenario where only two such parameters are free to vary: the coupling of 
the Higgs to $W$ and $Z$ vector bosons~($a$), and the coupling to fermions~($c$). Some of the simplest composite Higgs theories
in fact fall into this class, and we have reported explicit results for two benchmark models: a model a with universal rescaling of the Higgs
couplings (such as the MCHM4 and minimal conformal TC), and the MCHM5 model.

A fully consistent use of the current data to constrain the Higgs couplings in eq.~(\ref{eq:genL})
requires two important pieces of information to be reported by the experimental collaborations:
\begin{enumerate}
\item The likelihood for each channel as a function of the signal strength modifier $\mu$
\item The cut efficiencies for each channel and Higgs production  mode
\end{enumerate}
Unfortunately this information is not in general provided by ATLAS and CMS. We have however shown that
the body of LHC results published on the SM Higgs searches  is sufficient to allow one to derive an accurate estimate of the constraint
in a more general theory. In particular, we have designed a method to reconstruct the likelihood of each channel once given the
expected and observed 95\% CL limits on $\mu$. This technique  becomes rigorous in the asymptotic limit of large 
number of counts, and improves on more empirical
recipes used in the literature such as combining the limits in inverse quadrature. It has the further advantage of allowing a
best fit analysis in the case where a significant excess is observed compared to the pure background expectation.

By using our method we have derived the 95\%~CL limits implied by the full 2011 data set of CMS on $a$ and $c$, as well on the
parameter $\xi = (v/f)^2$ of the composite Higgs models MCHM4 and MCHM5. The results are shown in 
Figs.~\ref{fig:ac95},~\ref{fig:MCHM4},~\ref{fig:MCHM5}.
We have also performed a best fit analysis of the anomalous point at $m_h =125\,$GeV, for which both CMS and ATLAS have observed
a surplus of events in various channels, assuming the excess is due to the presence of the Higgs.
The resulting probability contours are reported in the plots of Fig.~\ref{fig:125} for  CMS and ATLAS respectively. The CMS data
seem to prefer a solution with negative $c$, for which the $\gamma\gamma$ decay rate is enhanced while the $WW$ and $ZZ$ rates
are close to the SM Higgs prediction. On the other hand, the large excess of ATLAS both in $\gamma\gamma$ and $ZZ$
seems to point to values $a >1$.
Although the emerging picture is not yet coherent, there are some conclusions which can be
already drawn from our analysis.

Perhaps the most important conclusion is that exclusive as opposed to inclusive searches are much more powerful to extract information
on the Higgs couplings, especially when the nature of the latter is non-standard. We have demonstrated that  this enhanced sensitivity is already
evident in the $\gamma\gamma$ and $WW$ channels when comparing the exclusive searches performed by CMS with the inclusive
ones carried out by ATLAS. Also, our analysis shows that a broader, model-independent interpretation of the Higgs searches can be 
performed easily and it should be the starting point to report future results. 

The explorative analysis performed in this work makes use of  all  data which is readily available in each channel and gives robust estimates of the limits currently imposed by the LHC searches on the couplings $a$, $c$. 
It  cannot be considered, however, as a substitute of the full, exact analysis which can be carried out only through  use of the complete experimental information.  We hope that such a full model-independent analysis  will be performed in the future
by the ATLAS and CMS collaborations.

\section*{Acknowledgments}
 
We thank Daniele Del Re and Shahram Rahatlou for participation in the early stages of this work and for many enlightening 
discussions and suggestions. We are especially indebted to  Emanuele Di Marco for patiently explaining to us the details of the $WW$ 
analysis of CMS, and for various important discussions and suggestions.
It is also a pleasure 
to thank Giulio D'Agostini, Vittorio Del Duca, Evan Friis, Christophe Grojean, Barbara Mele and Riccardo Rattazzi for stimulating discussions.  J.G. is grateful to the theory division  at FNAL for their hospitality during the completion of this project.  
The work of R.C. was partly supported by the ERC Advanced Grant No.~267985
{\em Electroweak Symmetry Breaking, Flavour and ÊDark Matter: One Solution for Three Mysteries (DaMeSyFla)}.


\begin{thebibliography}{9}

\bibitem{compositeHiggs}
  D.~B.~Kaplan and H.~Georgi,
  Phys.\ Lett.\  B {\bf 136} (1984) 183.
S.~Dimopoulos and J.~Preskill,
  Nucl.\ Phys.\  B {\bf 199}, 206 (1982).
T.~Banks,
  Nucl.\ Phys.\  B {\bf 243}, 125 (1984).
D.~B.~Kaplan, H.~Georgi and S.~Dimopoulos,
  Phys.\ Lett.\  B {\bf 136}, 187 (1984).
H.~Georgi, D.~B.~Kaplan and P.~Galison,
  Phys.\ Lett.\  B {\bf 143}, 152 (1984).
H.~Georgi and D.~B.~Kaplan,
  Phys.\ Lett.\  B {\bf 145}, 216 (1984).
M.~J.~Dugan, H.~Georgi and D.~B.~Kaplan,
  Nucl.\ Phys.\  B {\bf 254}, 299 (1985).



\bibitem{HiggsBounds}
  P.~Bechtle, O.~Brein, S.~Heinemeyer, G.~Weiglein and K.~E.~Williams,
  Comput.\ Phys.\ Commun.\  {\bf 181} (2010) 138
  [arXiv:0811.4169 [hep-ph]];
  Comput.\ Phys.\ Commun.\  {\bf 182} (2011) 2605
  [arXiv:1102.1898 [hep-ph]];


\bibitem{duhrssen}
M. Duhrssen, 
``Prospects for the measurement of Higgs boson coupling parameters in the mass range from $110-190$\,GeV$/c^2$'',
ATL-PHYS-2003-030.

\bibitem{Duhrssen:2004cv}
  M.~Duhrssen, S.~Heinemeyer, H.~Logan, D.~Rainwater, G.~Weiglein and D.~Zeppenfeld,
  Phys.\ Rev.\  D {\bf 70} (2004) 113009
  [arXiv:hep-ph/0406323].

\bibitem{Lafaye:2009vr}
  R.~Lafaye, T.~Plehn, M.~Rauch, D.~Zerwas and M.~Duhrssen,
  JHEP {\bf 0908} (2009) 009
  [arXiv:0904.3866 [hep-ph]].

\bibitem{Bonnet:2011yx}
  F.~Bonnet, M.~B.~Gavela, T.~Ota and W.~Winter,
  arXiv:1105.5140 [hep-ph].



\bibitem{grojean}
  J.~R.~Espinosa, C.~Grojean and M.~Muhlleitner,
  JHEP {\bf 1005} (2010) 065
  [arXiv:1003.3251 [hep-ph]];
  arXiv:1202.1286 [hep-ph].
  
  \bibitem{Bock:2010nz}
  S.~Bock, R.~Lafaye, T.~Plehn, M.~Rauch, D.~Zerwas and P.~M.~Zerwas,
  Phys.\ Lett.\ B {\bf 694} (2010) 44
  [arXiv:1007.2645 [hep-ph]].


\bibitem{Contino:2010mh}
  R.~Contino, C.~Grojean, M.~Moretti, F.~Piccinini and R.~Rattazzi,
  JHEP {\bf 1005} (2010) 089
  [arXiv:1002.1011 [hep-ph]].



\bibitem{Agashe:2004rs}
  K.~Agashe, R.~Contino and A.~Pomarol,
  Nucl.\ Phys.\  B {\bf 719} (2005) 165
  [arXiv:hep-ph/0412089].

\bibitem{Contino:2006qr}
  R.~Contino, L.~Da Rold and A.~Pomarol,
  Phys.\ Rev.\  D {\bf 75} (2007) 055014
  [arXiv:hep-ph/0612048].

\bibitem{Galloway:2010bp}
  J.~Galloway, J.~A.~Evans, M.~A.~Luty and R.~A.~Tacchi,
  JHEP {\bf 1010} (2010) 086
  [arXiv:1001.1361 [hep-ph]].


\bibitem{Giudice:2007fh}
  G.~F.~Giudice, C.~Grojean, A.~Pomarol and R.~Rattazzi,
  JHEP {\bf 0706} (2007) 045
  [arXiv:hep-ph/0703164].


\bibitem{dilaton}
  E.~Halyo,
  Mod.\ Phys.\ Lett.\  A {\bf 8} (1993) 275.
  W.~D.~Goldberger, B.~Grinstein and W.~Skiba,
  Phys.\ Rev.\ Lett.\  {\bf 100} (2008) 111802
  [arXiv:0708.1463 [hep-ph]];
  L.~Vecchi,
  Phys.\ Rev.\ D {\bf 82} (2010) 076009
  [arXiv:1002.1721 [hep-ph]];
  B.~A.~Campbell, J.~Ellis and K.~A.~Olive,
  arXiv:1111.4495 [hep-ph].

\bibitem{Alonso:2012px}
  R.~Alonso, M.~B.~Gavela, L.~Merlo, S.~Rigolin, J.~Yepes and ,
  arXiv:1212.3305 [hep-ph].

\bibitem{Low:2009di}
  I.~Low, R.~Rattazzi and A.~Vichi,
  JHEP {\bf 1004} (2010) 126
  [arXiv:0907.5413 [hep-ph]].

\bibitem{Azatov:2011qy}
  A.~Azatov and J.~Galloway,
  arXiv:1110.5646 [hep-ph].

\bibitem{Peskin:1991sw}
  M.~E.~Peskin and T.~Takeuchi,
  Phys.\ Rev.\ D {\bf 46} (1992) 381.

\bibitem{Barbieri:2004qk}
  R.~Barbieri, A.~Pomarol, R.~Rattazzi and A.~Strumia,
  Nucl.\ Phys.\ B {\bf 703} (2004) 127
  [hep-ph/0405040].

\bibitem{fit}
Alessandro Strumia, private communication. The data used in the fit are those of LEP1 (see Table 2 of Ref.~\cite{Barbieri:2004qk}), and those from 
Atomic Parity Violation (APV) (see Ref.~\cite{Barbieri:2004qk}, Table 3).  See also:
K.~Agashe and R.~Contino,
  Nucl.\ Phys.\ B {\bf 742} (2006) 59
  [hep-ph/0510164].

\bibitem{LHCsummer}
ATLAS and CMS Collaborations and LHC Higgs Combination Group, \textit{Procedure for the LHC Higgs boson search combination in Summer 2011}, 
CMS-NOTE-2011/005; ATL-PHYS-PUB-2011-11 (2011).

\bibitem{D'Agostini:2003nk}
  G.~D'Agostini,
  ``Bayesian reasoning in data analysis: A critical introduction,''
  New Jersey, USA: World Scientific (2003) 329 p


\bibitem{CMS:2012tw}
  CMS~Collaboration,
  ``Search for the standard model Higgs boson decaying into two photons in pp collisions at sqrt(s)=7 TeV,''
  arXiv:1202.1487 [hep-ex].
  
\bibitem{book}
  G.~Cowan,
  ``Statistical data analysis,"
  Oxford, UK: Clarendon (1998) 197 p.


\bibitem{CMS:combo}
  CMS~Collaboration,
  ``Combined results of searches for the standard model Higgs boson in pp collisions at sqrt(s) = 7 TeV''
  arXiv:1202.1488 [hep-ex].

\bibitem{CMS:comboPAS}
  CMS~Collaboration,
  ``Combination of CMS searches for a Standard Model Higgs boson''
  CMS-PAS HIG-11-032

\bibitem{Chatrchyan:2012ty}
  S.~Chatrchyan {\it et al.}  [CMS Collaboration],
  ``Search for the standard model Higgs boson decaying to a W pair in the fully leptonic final state in pp collisions at sqrt(s) = 7 TeV,''
  arXiv:1202.1489 [hep-ex].

\bibitem{WWtwiki} 
  ``Search for the Higgs Boson in the Fully Leptonic $W^+W^-$ Final State''
  CMS-PAS-HIG-11-024; {\tt https://twiki.cern.ch/twiki/bin/view/CMSPublic/Hig11024TWiki}.




\bibitem{ATLAS:combo}
  ATLAS~Collaboration,
  ``Combined search for the Standard Model Higgs boson using up to 4.9 fb-1 of pp collision data at sqrt(s) = 7 TeV with the ATLAS detector at the LHC,''
  arXiv:1202.1408 [hep-ex].

\bibitem{Falkowski:2012vh}
  A.~Falkowski, S.~Rychkov and A.~Urbano,
  arXiv:1202.1532 [hep-ph].
  
\bibitem{rattazzi}
R.~Rattazzi, talk given at the ETH workshop ``Higgs Searches confronts theory'',  9-11 January 2012, Zurich

\bibitem{SCTC}
  A.~Azatov, J.~Galloway and M.~A.~Luty,
  Phys.\ Rev.\ Lett.\  {\bf 108}, 041802 (2012)
  [arXiv:1106.3346 [hep-ph]];
  A.~Azatov, J.~Galloway and M.~A.~Luty,
  Phys.\ Rev.\ D {\bf 85}, 015018 (2012)
  [arXiv:1106.4815 [hep-ph]].


\bibitem{Akeroyd:1998ui}
ÊA.~G.~Akeroyd,
Ê
ÊJ.\ Phys.\ G G {\bf 24}, 1983 (1998)
Ê[hep-ph/9803324].
Ê

\bibitem{Ferreira:2011aa}
ÊP.~M.~Ferreira, R.~Santos, M.~Sher and J.~P.~Silva,
Ê
ÊarXiv:1112.3277 [hep-ph].
Ê

\bibitem{Gabrielli:2012yz}
  E.~Gabrielli, B.~Mele and M.~Raidal,
  arXiv:1202.1796 [hep-ph].

\bibitem{us}
  A.~Azatov, R.~Contino, D.~Del Re, J.~Galloway, M.~Grassi and S.~Rahatlou,
ÊÊarXiv:1204.4817 [hep-ph].
ÊÊ












\end{thebibliography}
\end{document}